\documentclass[apj]{emulateapj}
\usepackage{graphicx}
\usepackage[toc,page]{appendix}
\usepackage{natbib}
\usepackage{color}
\usepackage{threeparttable}
\usepackage{hyperref}
\usepackage{breakurl}
\begin{document}
\slugcomment{Submitted to the Astrophysical Journal Letters}
\title{The Star-formation Mass Sequence out to $z=2.5$}
\email{katherine.whitaker@yale.edu}
\author{Katherine E. Whitaker\altaffilmark{1}, 
Pieter G. van Dokkum\altaffilmark{1}, Gabriel Brammer\altaffilmark{2}, Marijn Franx\altaffilmark{3}}
\altaffiltext{1}{Department of Astronomy, Yale University, New Haven, CT 06511}
\altaffiltext{2}{European Southern Observatory, Alonso de C\'{o}rdova 3107, Casilla 19001, Vitacura, Santiago, Chile}
\altaffiltext{3}{Sterrewacht Leiden, Leiden University, NL-2300 RA Leiden, The Netherlands}

\shortauthors{Whitaker et al.}
\shorttitle{The Star-formation Mass Sequence out to $z=2.5$}

\begin{abstract}
We study the star formation rate (SFR) - stellar mass (M$_{\star}$) relation in a self-consistent manner from $0<z<2.5$
with a sample of galaxies selected from the NEWFIRM Medium-Band Survey.  
We find a significant non-linear slope of the relation, 
$\mathrm{SFR}\propto\mathrm{M}_{\star}^{0.6}$, and a 
constant observed scatter of 0.34 dex, independent of redshift and M$_{\star}$.
However, if we select only blue galaxies we find a linear relation $\mathrm{SFR}\propto\mathrm{M}_{\star}$,
similar to previous results at $z=0$ by \citet{Peng10}.
This selection excludes red, dusty, star-forming galaxies with higher masses,
which brings down the slope.
By selecting on $\mathrm{L_{IR}/L_{UV}}$ (a proxy for dust obscuration) and the
rest-frame $U$--$V$ colors, we show that star-forming galaxies fall in three distinct regions of the
log(SFR)-log(M$_{\star}$) plane: 1) actively star-forming 
galaxies with ``normal'' dust obscuration and associated colors
(54\% for $\mathrm{log(M_{\star})>10}$ at $1<z<1.5$),
2) red star-forming galaxies with low levels of dust obscuration
and low specific SFRs (11\%), and 3) dusty, blue star-forming galaxies with 
high specific SFRs (7\%).  The remaining 28\% comprises quiescent galaxies.  
Galaxies on the ``normal'' star formation sequence show strong
trends of increasing dust attenuation with stellar mass and a decreasing specific SFR, with an observed scatter of 
0.25 dex (0.17 dex intrinsic scatter).
The dusty, blue galaxies reside in the upper envelope of the star formation sequence
with remarkably similar spectral shapes at all masses, suggesting that the same physical process is dominating
the stellar light.  The red, low-dust star-forming galaxies may be in the process of shutting 
off and migrating to the quiescent population.  
\end{abstract}


\section{Introduction}
\label{sec:intro}

Galaxies show a strong correlation between their star formation rate (SFR)
and stellar mass (M$_{\star}$) from $z=0$ to the earliest observed epoch, 
$z=7$
\citep[e.g.,][]{Brinchmann04,Noeske07a,Elbaz07,Daddi07,Pannella09,Magdis10,Gonzalez10}.
On average, galaxies on this ``star formation sequence'' 
were forming stars at much higher rates in the distant universe 
relative to today \citep[e.g.,][]{Madau96}; for a given mass, the SFR 
has been decreasing at a steady rate by a factor of $\sim30$ from $z\sim2$ to $z=0$ \citep{Daddi07}, 
although it appears to be roughly constant from $z\sim7$ to 
$z\sim2$ \citep{Gonzalez10}.  
The star formation sequence is observed to have a roughly
constant scatter out to $z\sim1$ \citep[e.g.,][]{Noeske07a}.

Generally, star formation is thought to be regulated by the balance between the 
rate at which cold gas is accreted onto the galaxy and feedback 
\citep[e.g.,][]{Dutton10, Bouche10}, whereas the relation between gas
surface density and SFR surface density does not appear to change, at least out to the highest redshifts 
accessible to molecular gas studies \citep{Daddi10,Tacconi10}.  
The star formation sequence may be a natural consequence of `cold mode accretion' \citep[e.g.,][]{Birnboim03},
as the SFR is approximately a steady function of time and yields a relatively tight relationship 
between SFR and M$_{\star}$.  Feedback may effect the slope of the star formation sequence,
whereas
the evolution of the normalization is thought to result from evolution 
in gas densities with redshift.  The scatter in the star formation sequence 
may reflect variations in the gas accretion history,
and is predicted to be insensitive to stellar mass, redshift and feedback efficiencies \citep{Dutton10}.

With a wealth of multi-wavelength data,                       
including ultraviolet (UV) to near-infrared (NIR) photometric surveys, H$\alpha$ spectroscopic surveys, and
mid- to far-infrared photometry from \emph{Spitzer}/MIPS at 24$\mu$m,\emph{Herschel}/PACS 70--160$\mu$m,
and \emph{Herschel}/SPIRE 250--500$\mu$m,
much work has been done to calibrate different SFR indicators over broad redshift
ranges \citep[e.g.,][]{Muzzin10, Hwang10, Wuyts11a, Wuyts11b, Reddy12}.
Uncertainties in the star formation sequence may now be dominated by 
selection effects and observational biases.
Selection effects can be important, as it is well known that a significant fraction of 
galaxies have very low SFRs, well below the star formation sequence.  Furthermore,
there exists a population of dusty star-forming galaxies with similar rest-frame optical 
colors to these quiescent
galaxies~\citep[e.g.,][]{Brammer09}. The selection of 
star-forming galaxies and the treatment of
quiescent galaxies can influence the measured relation and its scatter.  

\begin{figure*}[t]
\leavevmode
\centering
\includegraphics[width=0.87\linewidth]{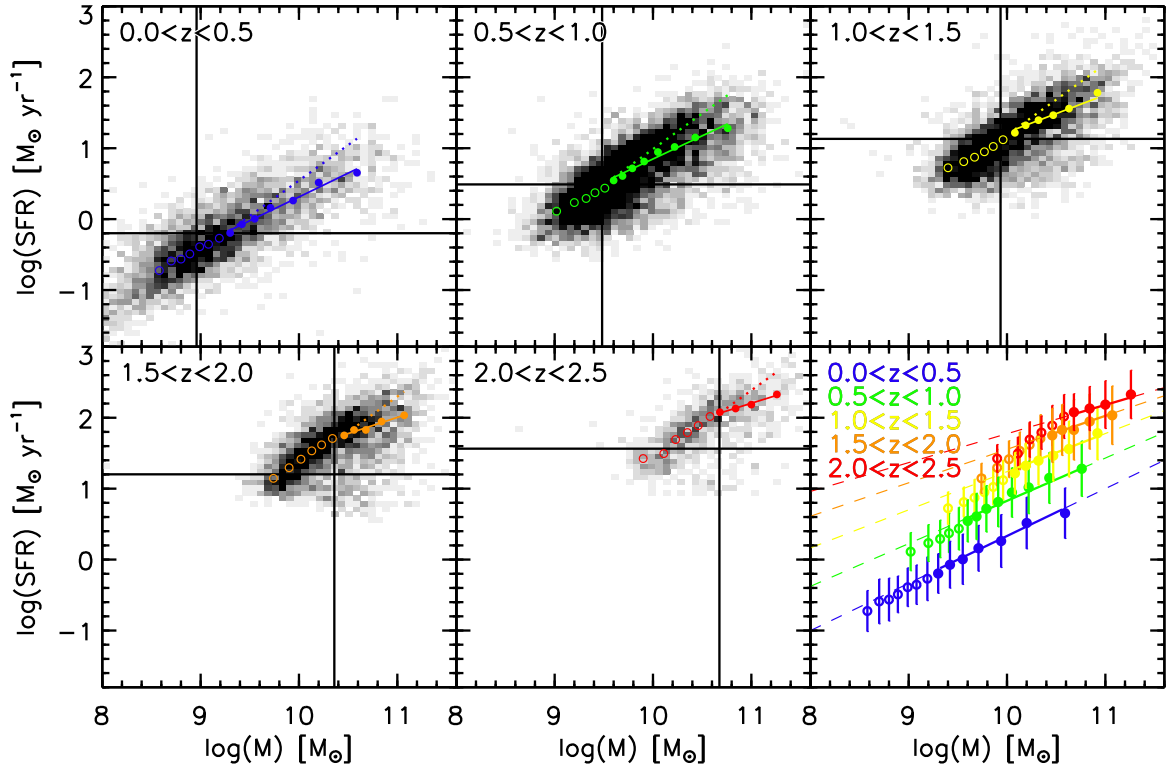}
\caption{The SFR-mass sequence for star-forming galaxies has a non-linear slope out
to $z=2.5$ (dotted line is linear).  
The running medians and scatter are color-coded by redshift, with a power law fit above the mass
and SFR completeness limits (solid lines in bottom, right panel).  }
\label{fig:sfr}
\end{figure*}

With the accurate photometric redshifts and photometry of the NEWFIRM Medium-Band Survey 
\citep[NMBS; ][]{Whitaker11}, we can now extend studies of the star formation sequence out to $z=2.5$ for the 
largest mass-complete sample of galaxies to date.  In this letter, we study the evolution
of the log(SFR)-log(M$_{\star}$) relation from $0<z<2.5$,
selecting star-forming galaxies in a way that is independent from the SFR indicator.
We show that star-forming galaxies with 
different colors and $\mathrm{L_{IR}/L_{UV}}$ ratios occupy different 
regions of the log(SFR)-log(M$_{\star}$) plane. 

We assume a $\Lambda$CDM cosmology with $\Omega_{M}$=0.3, $\Omega_{\Lambda}$=0.7, 
and $H_{0}$=70 km s$^{-1}$ Mpc$^{-1}$ throughout the paper.  

\section{Data and Sample Selection}
\label{sec:data}

Our sample of galaxies is drawn from the NMBS \citep{Whitaker11}.
This survey employs a new technique of using five medium-bandwidth NIR filters to sample the
Balmer/4000\AA\ break from $1.5<z<3.5$ at a higher resolution than the standard
broadband NIR filters.  The combination of the medium-band NIR images with
deep optical medium and broadband photometry and IRAC imaging over 0.4 deg$^{2}$
results in accurate photometric redshifts
($\Delta z/(1+z)\lesssim 2\%$), rest-frame colors and stellar population
parameters.  
The SFRs presented in this paper are based in part on \emph{Spitzer}-MIPS fluxes at 24$\mu$m that are derived
from the S-COSMOS \citep{Sanders07} and 
FIDEL\footnote{\url{http://irsa.ipac.caltech.edu/data/SPITZER/FIDEL/}} surveys.
A comprehensive overview of the survey can be found in \citet{Whitaker11}.  
The stellar masses used in this work
are derived using FAST \citep{Kriek09a}, with \citet{BC03} models 
that assume a \citet{Chabrier} initial mass function (IMF), solar metallicity, exponentially declining star formation
histories and dust extinction following the \citet{Calzetti00} extinction law.

The SFRs are determined by adding the UV and IR emission, 
$\mathrm{SFR_{UV+IR}}=0.98\times10^{-10}(L_{\mathrm{IR}}+3.3L_{2800})$ 
\citep{Kennicutt98}, adapted for the Kroupa IMF by \citet{Franx08},
accounting for the unobscured and obscured star 
formation, respectively.  We adopt a luminosity-independent conversion from the observed 24$\mu$m flux
to the total IR luminosity ($L_{\mathrm{IR}}\equiv L$(8--1000$\mu$m)), based on a single template that is
the log average of \citet{DH02} templates with $1<\alpha<2.5$, following
\citet{Wuyts08,Franx08,Muzzin10}, and in good median agreement with recent \emph{Herschel}/PACS measurements
by \citet{Wuyts11a}. 
The luminosities at 2800\AA\ ($L_{\mathrm{2800}}$) are derived
directly from the best-fit template to the observed photometry, 
using the same methodology as the rest-frame colors \citep[see][]{Brammer11}.

With accurate rest-frame colors, it is possible to isolate ``clean'' samples of star-forming and quiescent galaxies 
using two rest-frame colors out to high redshifts \citep{Labbe05,Wuyts07,Williams09,Ilbert09,Brammer11,Whitaker11}.  
The quiescent galaxies have strong Balmer/4000\AA\ breaks, characterized by red $U$--$V$ colors and bluer
$V$--$J$ colors relative to dusty star-forming galaxies at the same $U$--$V$ color.

\citet{Whitaker11} demonstrated that there is a clear delineation between star-forming and quiescent galaxies
with the NMBS data set.
Quiescent galaxies are identified using the criteria $U-V > 0.8\times(V-J)+0.7$, 
$U-V>1.3$ and $V-J<1.5$, and they are excluded from the bulk of this analysis.
The sample of star-forming galaxies is selected independent of the 
SFR indicator and stellar population synthesis model parameters, enabling an unbiased
measurement of the star formation sequence.

\section{The Star Formation Sequence}
\label{sec:sfr}

Figure~\ref{fig:sfr} shows the star formation sequence, log(M$_{\star}$)--log(SFR), 
in five redshift bins out to $z=2.5$.
The greyscale represents the density of points for star-forming galaxies selected in Section~\ref{sec:data},
with the running median and biweight scatter color-coded by redshift.    
The mass completeness limits are estimated from the 90\% point-source completeness limits derived from
the unmasked simulations by \citet{Whitaker11}.
The SFR completeness limits correspond to the 3$\sigma$ 24$\mu$m
detection limit (17.6 $\mu$Jy) at the highest redshift of each bin.
All 24$\mu$m detections $<1\sigma$ are replaced with the $1\sigma$ upper limit,
resulting in a flattened
tail of the log(SFR)-log(M$_{\star}$) relation at low M$_{\star}$, where the samples are incomplete.  

\subsection{Quantifying the Star Formation Sequence}
\label{sec:quantify_sfr}

The running medians and dispersions are measured for all star-forming galaxies, and those 
above the mass and SFR completeness limits are indicated with filled symbols in Figure~\ref{fig:sfr} and fit 
with a power law,
\begin{equation}
\log(\mathrm{SFR})=\alpha(z)(\log\mathrm{M_{\star}}-10.5)+\beta(z)
\label{eq:ms}
\end{equation}

We find that the slope $\alpha$ gradually evolves with redshift, while the normalization $\beta$
has a stronger evolution:
\begin{eqnarray}
  \alpha(z) &=& 0.70-0.13z\\
  \beta(z) &=& 0.38+1.14z-0.19z^2
\end{eqnarray}

Previous studies have found hints of a similar trend for the slope to
flatten toward $z\sim1$ or high M$_{\star}$ \citep[e.g.,][]{Noeske07a,Karim11,Wuyts12}.
Although the slope is consistent with a gradual evolution toward shallower values,
we note that the highest redshift bins are only 
for galaxies with stellar masses
$>10^{10.7}$ M$_{\odot}$, due to incompleteness at lower masses.  
We will address this issue in Section~\ref{sec:dust}.

The average SFR of star-forming galaxies steeply declines from $z=2.5$
to the present day, changing by roughly 0.2 dex Gyr$^{-1}$, confirming the results
of previous studies \citep[e.g.][etc.]{Noeske07a, Elbaz07}.
The normalization $\beta$ decreases by a factor of 4 from $z=2$ to $z=0$, while the global
SFR density has decreased a factor of 6 over the same interval \citep{Hopkins06}.
We note that $\beta(z)$ is the evolution in the specific SFR
($\mathrm{sSFR}\equiv\mathrm{SFR/M}_{\star}$) at log(M$_{\star})=10.5$
\citep[consistent with][etc]{Elbaz07,Daddi08,Pannella09,Damen09,Magdis10}.

Interestingly, the scatter in the star formation sequence is observed to be roughly 
constant at $\sim0.34$ dex, both with redshift and stellar mass, consistent with the idea
that it reflects variations in the gas accretion history.
The observed scatter agrees with results from \citet{Noeske07a} at $z<1$,
although we note that there are hints of a lower scatter
for the most massive galaxies in our highest redshift bin.  We find that these results are robust against
changes in sample selection.  If we include
all galaxies (quiescent and star-forming) with $>1\sigma$ detections, the scatter
increases by 0.07--0.11 dex.  

\begin{figure}[t]
\leavevmode
\centering
\includegraphics[width=0.9\linewidth]{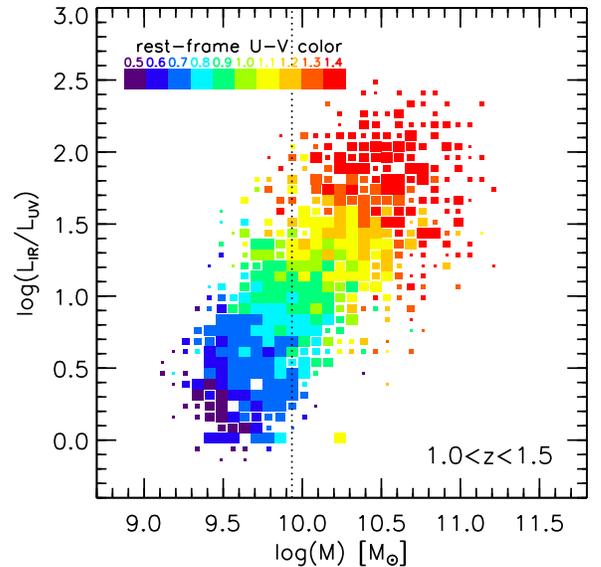}
\caption{More massive star-forming galaxies have higher log($\mathrm{L_{IR}/L_{UV}}$) ratios and
redder rest-frame $U$--$V$ colors, suggesting increasing amounts of dust attenuation. The dotted line is 
the mass-completeness limit.}
\label{fig:LIRLUV}
\end{figure}

\begin{figure*}[t]
\begin{minipage}[h]{0.33\linewidth}
\centering
\includegraphics[width=0.95\linewidth]{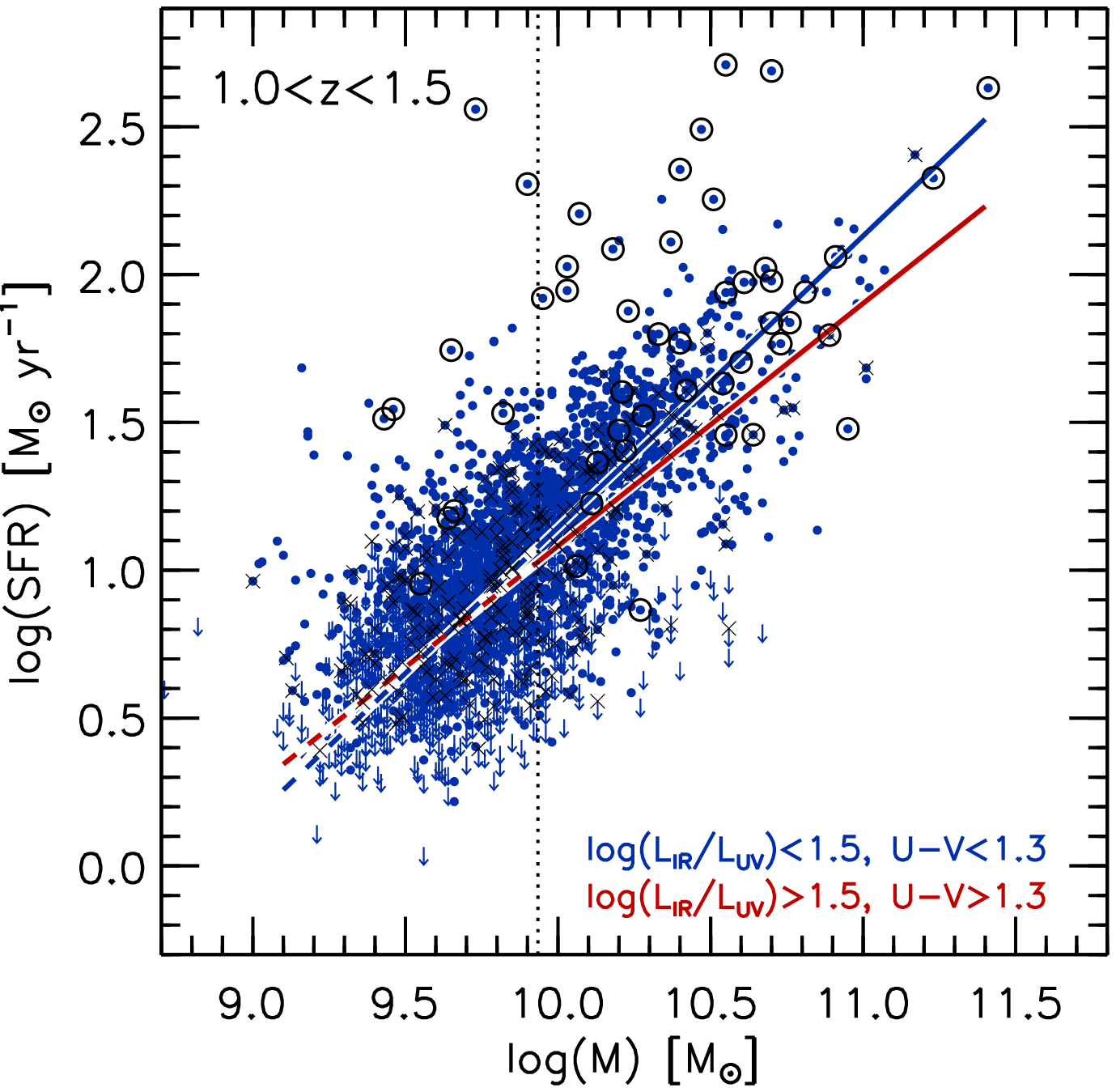}
\end{minipage}
\begin{minipage}[h]{0.33\linewidth}
\centering
\includegraphics[width=0.95\linewidth]{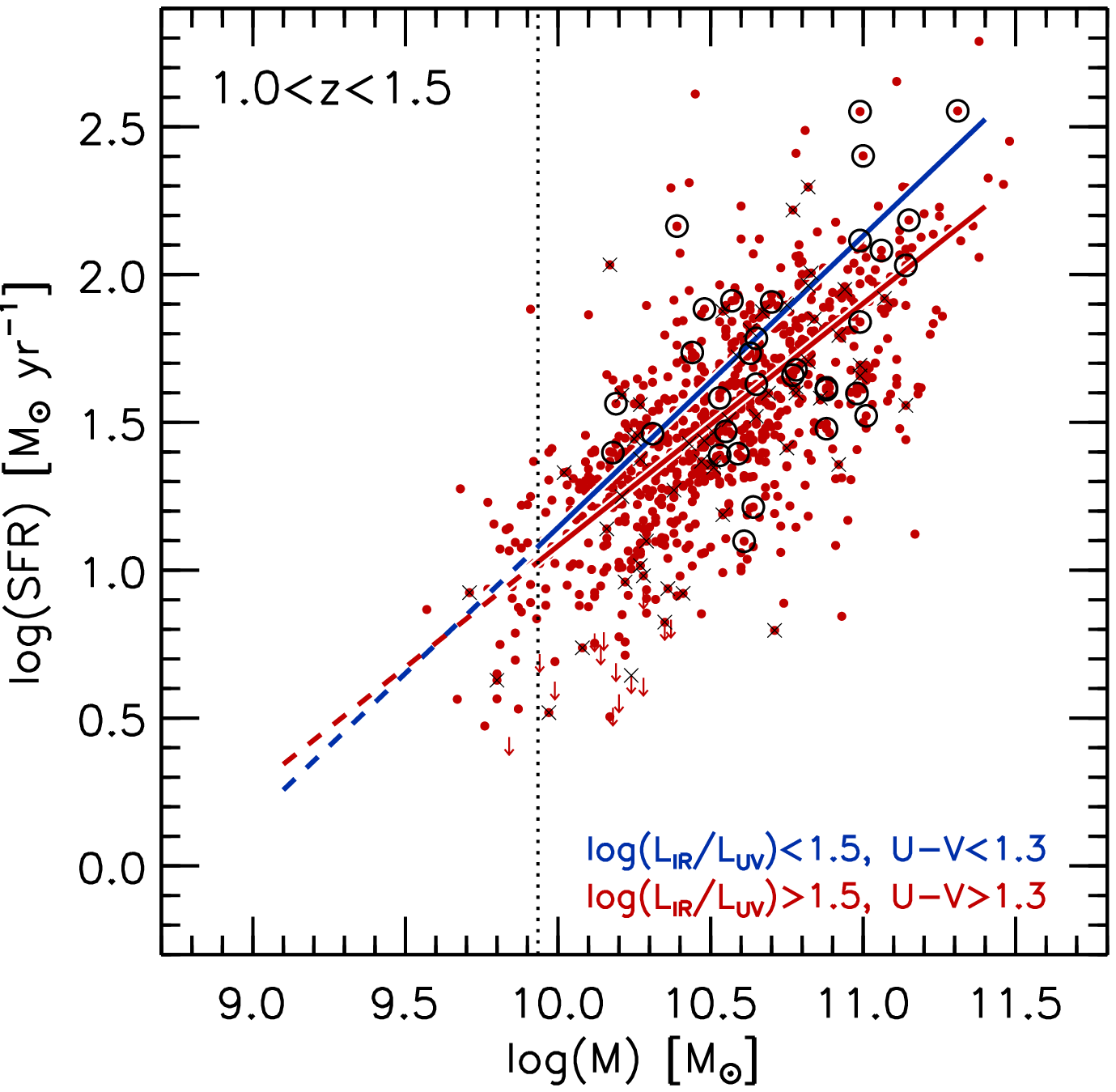}
\end{minipage}
\begin{minipage}[h]{0.33\linewidth}
\centering
\includegraphics[width=0.95\linewidth]{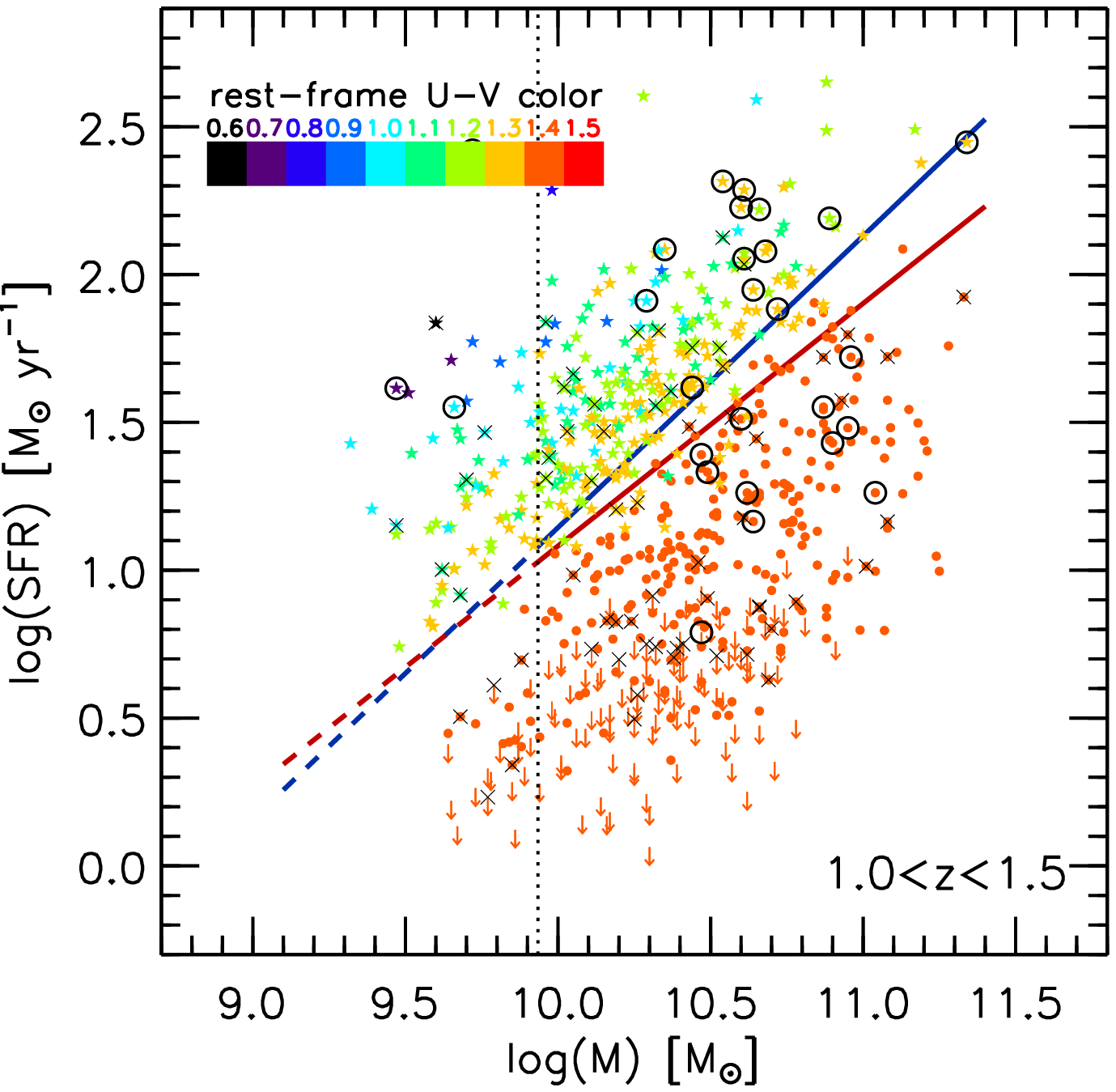}
\end{minipage}
\caption{The log(SFR)-log(M$_{\star}$) relation for blue star-forming galaxies with low dust has a best-fit
slope close to unity (left), whereas dusty, red star-forming galaxies follow a shallower slope (middle).
Although selected independent of their measured SFRs and M$_{\star}$, blue, dusty galaxies and
red, low-dust galaxies lie above and below the star formation sequence, respectively (right).
X-ray detections are indicated with a black open circles, and those galaxies without X-ray coverage 
are indicated with crosses. }
\label{fig:selection}
\end{figure*}

\subsection{The Effects of Dust Attenuation}
\label{sec:dust}

Previous studies have noted that dust attenuation is a strong function of stellar mass
\citep[e.g.,][]{Reddy06,Reddy10,Pannella09, Wuyts11b}, where the most massive galaxies are more highly obscured.
We find a similar trend with the NMBS data in Figure~\ref{fig:LIRLUV};
higher mass galaxies have larger $\mathrm{L_{IR}/L_{UV}}$ ratios.  The SFRs derived
from $L_{\mathrm{2800}}$ alone exhibit a flat trend with stellar mass (or even hints of less
UV flux toward higher M$_{\star}$), whereas SFRs derived from $L_{\mathrm{IR}}$
show a strong trend with M$_{\star}$.  The star formation-mass relation only materializes
when dust attenuation is properly taken into account.

From Figure~\ref{fig:LIRLUV}, we also see that the average rest-frame $U$--$V$ color of star-forming
galaxies becomes redder with increasing stellar mass (and increasing $\mathrm{L_{IR}/L_{UV}}$).  
The red colors may be attributed
to dust, although older stellar populations have similarly red colors.
It is notoriously difficult to differentiate between subtle age
and dust effects with a single color alone.  However, because
$\mathrm{L_{IR}}$ is independent of the rest-frame color measurements, we can
isolate galaxies with red colors predominantly due to dust.

The strong dependence of $\mathrm{L_{IR}/L_{UV}}$ and the rest-frame $U$--$V$ color
on stellar mass raises the question of whether the star formation sequence can be
``resolved'' into distinct populations of star-forming galaxies.  
We first consider the following two samples at $1.0<z<1.5$ (where we have a large number of 
galaxies at the high-mass end
while maintaining a modest mass-completeness level):
1) blue star-forming galaxies with little dust 
obscuration ($U-V<1.3$, log$(\mathrm{L_{IR}/L_{UV}})<1.5$), and 2) dusty red star-forming galaxies
($U-V>1.3$, log$(\mathrm{L_{IR}/L_{UV}})>1.5$).  

When we select blue star-forming galaxies,
similar to \citet{Peng10}, the slope of the star formation sequence is close to unity and
the observed scatter decreases to 0.25 dex (left panel of Figure~\ref{fig:selection}).  
However, dusty red star-forming galaxies are an increasing fraction of the
massive galaxy population toward $z=2$ \citep[e.g.,][]{Whitaker10}.
The middle panel of Figure~\ref{fig:selection} shows the star formation sequence for galaxies with 
red colors that we attribute to dust due to the high $\mathrm{L_{IR}/L_{UV}}$ ratios.  
This population of galaxies has a similarly small observed scatter of 0.25 dex, but a shallower slope of $\sim0.8$.

While it has been shown that higher mass galaxies tend to have lower
sSFRs \citep[e.g.,][]{Zheng07}, previous studies have not always distinguished between
actively star-forming and passive
stellar populations. Here, we find that 
\emph{among} actively star-forming galaxies, the star formation sequence for red dusty (high-mass) galaxies 
has a shallower slope, as compared to that for blue low-dust (low-mass) galaxies.
This result implies that both massive quiescent and star-forming galaxies have 
lower sSFRs and hence older ages.

\subsection{What causes galaxies to be outliers on the star formation sequence?}

We next consider the sample of star-forming galaxies with ``anomolous'' combinations 
of rest-frame color and $\mathrm{L_{IR}/L_{UV}}$ ratios to determine 
where these galaxies lie in the SFR-M$_{\star}$ plane.
We stress that we are not \emph{selecting} 
these galaxies to be outliers in the log(SFR)-log(M$_{\star}$) plane, 
rather to be unusual in their observed color and dust properties.
The galaxies with red rest-frame colors and low $\mathrm{L_{IR}/L_{UV}}$ ratios
fall in the lower envelope of the star formation sequence 
(right panel of Figure~\ref{fig:selection}).  These low sSFR galaxies 
may be in the process of shutting down star formation.  
Similarly, we identify galaxies that have blue colors but high $\mathrm{L_{IR}/L_{UV}}$ ratios.  
We find that dusty, blue galaxies have high sSFRs, predominantly residing in the upper envelope of the star 
formation sequence. This suggests that they are in a starburst phase 
and/or may have active galactic nuclei (AGN).

To search for signatures of AGN, we match the NMBS catalogs within a 1 arcsecond radius               
to the X-ray point source catalogs from the AEGIS-X survey \citep{Laird09}                          
and the Chandra COSMOS survey \citep{Elvis09}.  Those galaxies with counterparts in                 
the X-ray catalogs are indicated with open circles in Figure~\ref{fig:selection}.
We find that the fraction of galaxies with X-ray detections is $\sim4\%$ in all populations of star-forming
galaxies, implying that the central black holes may become active at any stage of star
formation \citep[e.g.,][]{Santini12}.
Interestingly, X-ray detected galaxies tend to be outliers in the log(SFR)-log(M$_{\star}$) plane (see
Figure~\ref{fig:selection}), probably because a fraction of the 24$\mu$m emission is not
associated with star formation, as assumed.
 
\begin{figure*}[t]
\leavevmode
\hspace{0.075\linewidth}
\begin{minipage}[h]{0.4\linewidth}
\centering
\includegraphics[width=0.94\linewidth]{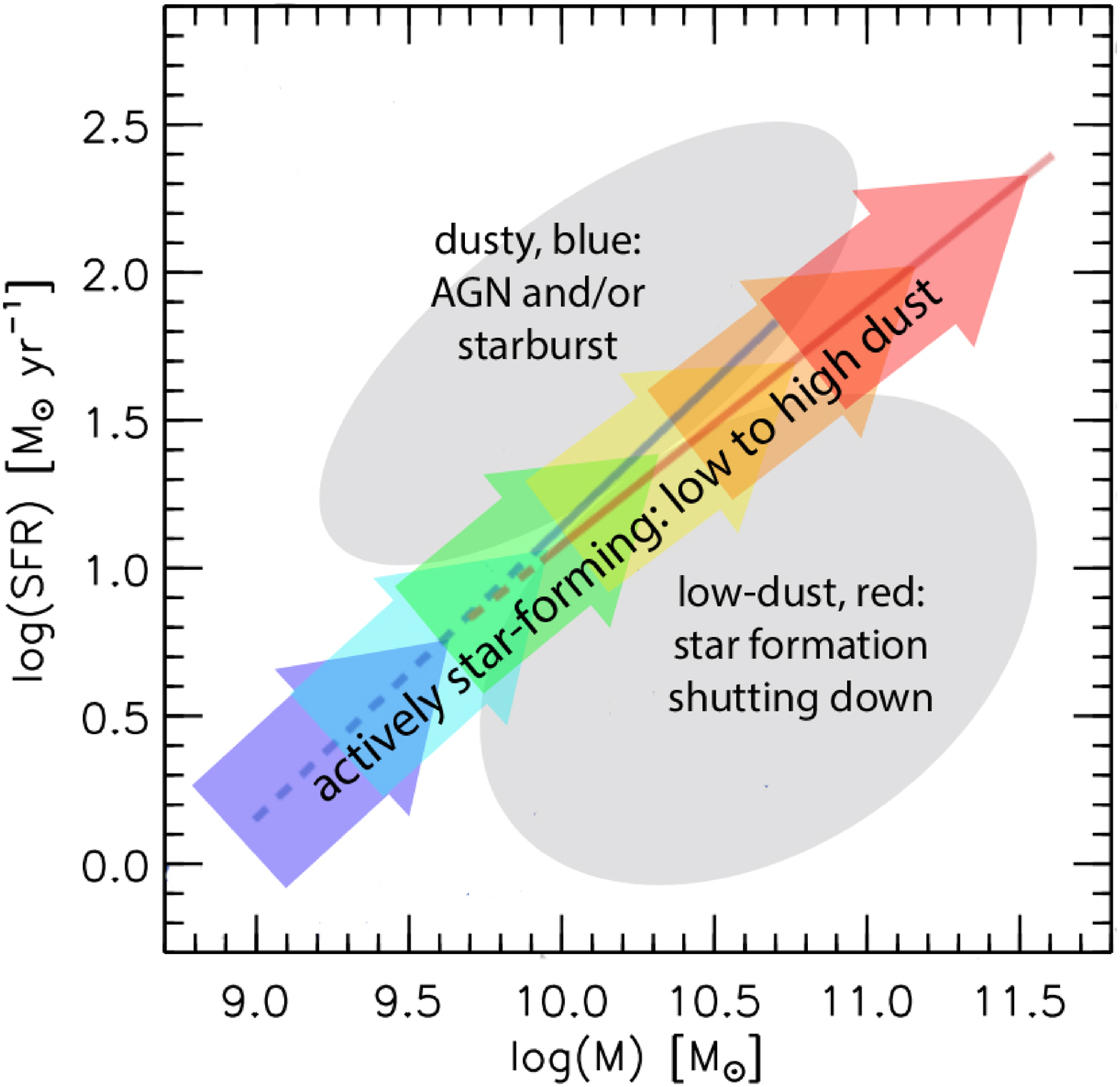}
\end{minipage}
\begin{minipage}[h]{0.4\linewidth}
\centering
\includegraphics[width=0.94\linewidth]{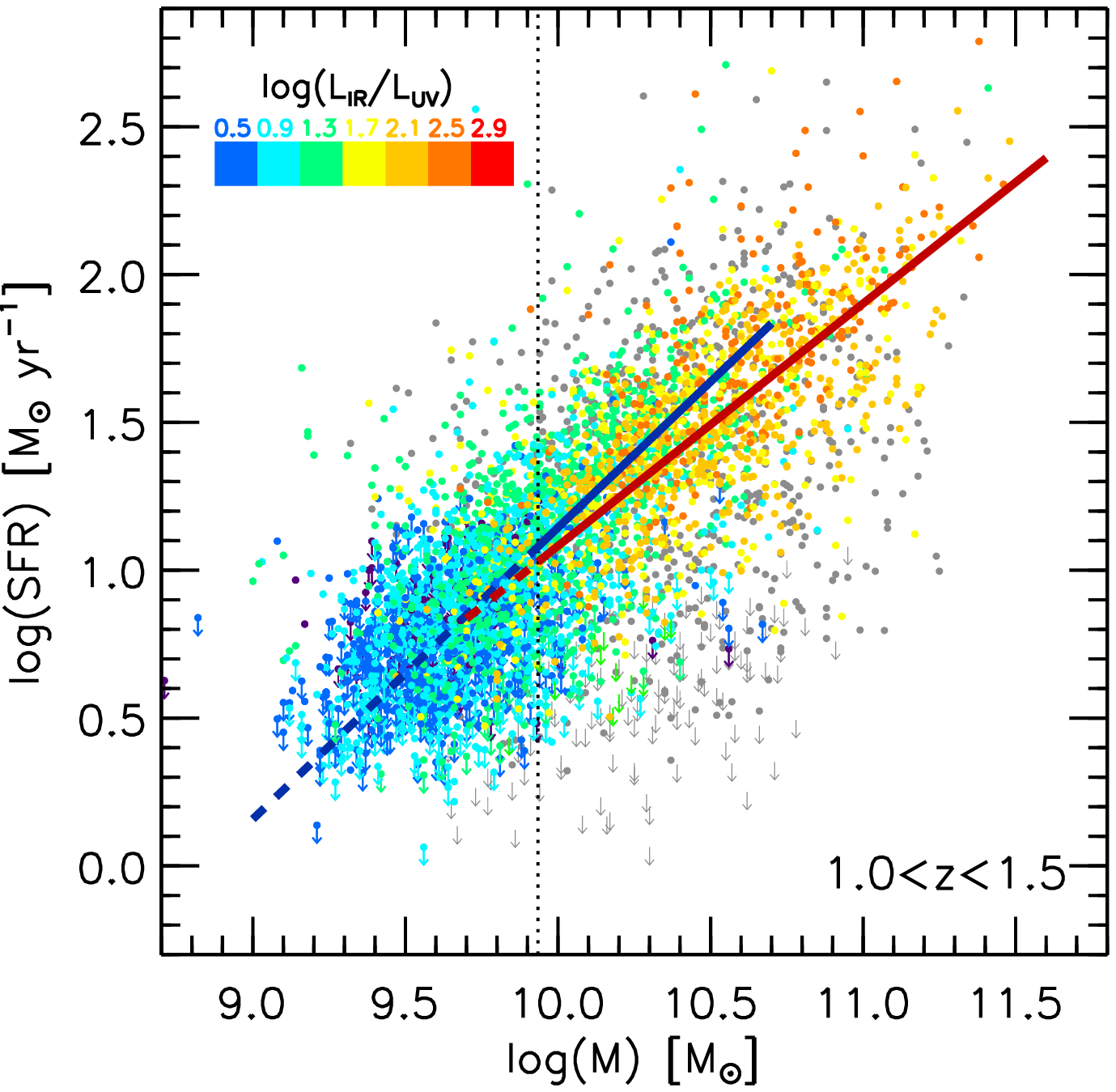}
\end{minipage}
\hspace{0.1\linewidth}
\caption{A cartoon (left) and observed (right) view of
how star-forming galaxies populate the log(SFR)-log(M$_{\star}$) plane, with
54\% of all galaxies residing on the ``normal'' star formation sequence, showing increasing amounts of dust (higher
$\mathrm{L_{IR}/L_{UV}}$ ratios) and lower sSFRs (shallower slope) towards higher stellar mass.
7\% of galaxies have blue colors and high $\mathrm{L_{IR}/L_{UV}}$ ratios, falling in the upper
envelope (grey). 11\% of galaxies have low $\mathrm{L_{IR}/L_{UV}}$ ratios and red colors,
populating the lower envelope (also grey).}
\label{fig:demo}
\end{figure*}

\section{Discussion}
\label{sec:discussion}

We find that the star formation sequence is not linear, with 
$\mathrm{SFR}\propto\mathrm{M}_{\star}^{0.6}$ and a constant observed scatter of 0.34 dex.
If we only select blue galaxies, however, we do find a linear relation, similar to \citet{Peng10}.  
This selection removes red, dusty star-forming galaxies at the high mass end, which have a shallower slope. 
We note that \citet{Pannella09} also find a slope close to unity at $z\sim2$, however this may be a result of the
``BzK'' selection.
Here, we are able to analyze the star formation sequence with a mass-complete 
sample of galaxies.

By accounting for the ratio of $\mathrm{L_{IR}/L_{UV}}$ (a proxy for dust) and the 
rest-frame $U$--$V$ colors of star-forming galaxies, we develop a simple picture
that describes how galaxies populate the log(SFR)-log(M$_{\star}$) plane.
We have identified four distinct populations: 1) quiescent galaxies (28\% for 
$\log(\mathrm{M}_{\star})>10$ at $1.0<z<1.5$), 2) actively star-forming galaxies with ``normal'' 
colors and associated $\mathrm{L_{IR}/L_{UV}}$ ratios (54\%), 3) red star-forming
galaxies with low $\mathrm{L_{IR}/L_{UV}}$ ratios and low sSFRs (11\%), and 
4) blue star-forming galaxies with high $\mathrm{L_{IR}/L_{UV}}$ ratios and sSFRs (7\%).

Among the galaxies that populate the ``normal'' star formation sequence, 
we see a continuous sequence of increasing levels of dust attenuation with increasing stellar mass
and a decreasing slope in the log(SFR)-log(M$_{\star}$) relation, implying decreasing sSFRs with stellar mass
(Figure~\ref{fig:demo}).  \citet{Wuyts11b} observed similar trends across the SFR-M$_{\star}$ plane,
where $\mathrm{L_{IR}/L_{UV}}$ increases along the sequence and at a given M$_{\star}$ towards high SFRs.
The dependence of sSFR on M$_{\star}$ appears to introduce a slight curvature to the star formation sequence.
A power-law fit to \emph{all} star-forming populations results in an observed scatter that is 0.09 dex larger than that
of the ``normal'' sequence, with a significantly shallower slope of $\sim0.6$.
We note that some models predict slopes that are too steep compared 
to the observations \citep[e.g.,][]{Bouche10,Dutton10}; it is possible that
discrepancies between the inferred SFRs may be alleviated if the stellar IMF is 
systematically weighted toward more high-mass star formation in rapidly star-forming galaxies \citep{Narayanan12}.

The observed scatter of the ``normal'' star formation sequence is 0.25 dex, which includes contributions 
from both random and systematic errors.
We estimate the average random scatter by
perturbing the 24$\mu$m photometry and photometric redshifts within the $1\sigma$ error bars for 100 realizations,
finding a small contribution of $\sim0.05$ dex.  Additionally, about 0.08 dex scatter is 
introduced because the average SFR evolves within the redshift bin.  Finally, uncertainties in the conversion 
from 24$\mu$m flux to $\mathrm{L_{IR}}$ introduce $\sim0.15$ dex scatter \citep{Marcillac06},
although we note that this value is quite uncertain \citep[e.g.,][]{Wuyts11a}. 
In total, we estimate that random and systematic errors introduce 0.18 dex scatter to the star formation
sequence, from which we estimate the intrinsic scatter to be 0.17 dex .  
Semi-analytic models predict an even smaller scatter ($0.12\pm0.1$ dex).  However, the model scatter
may be underestimated due to a simplified treatment of the halo mass accretion history.  Cosmological 
hydrodynamical simulations by \citet{Dekel09} find a scatter of up to 0.3 dex in the gas accretion rates
on to galaxies in $10^{12}$ M$_{\odot}$ haloes at $z=2.5$.  If this scatter translates linearly into 
scatter in the SFRs, this may reconcile the differences between the models and observed scatter.   

\begin{figure*}[t]
\leavevmode
\centering
\includegraphics[width=0.7\linewidth]{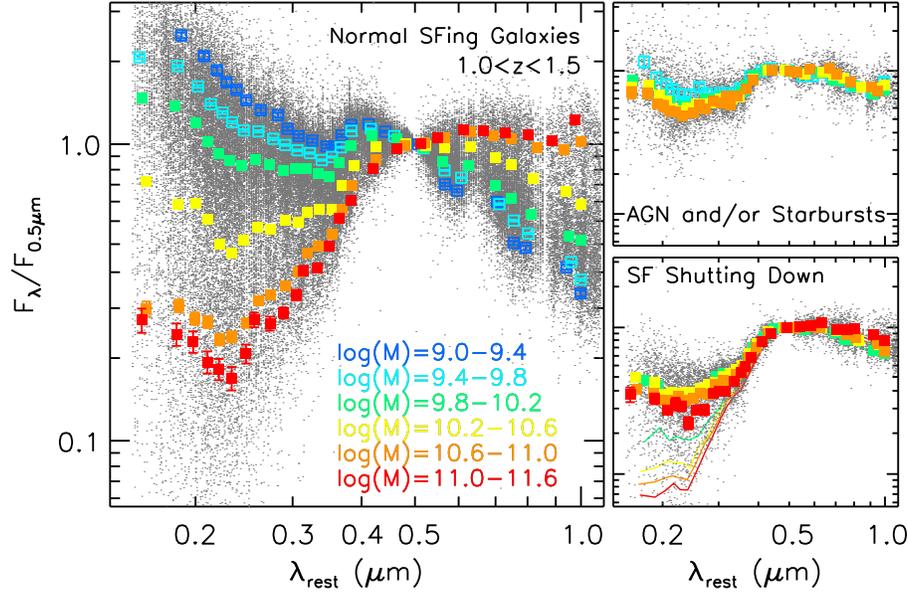}
\caption{The rest-frame composite SEDs of galaxies on the ``normal'' star formation sequence (left)
showing increasing levels of dust attenuation with stellar mass.  The spectral shape of dusty, blue
galaxies appears to be independent of stellar mass (upper right).  Galaxies in the process of shutting off
their star formation (bottom right) show larger amounts of rest-frame UV emission relative to quiescent galaxies at
the same stellar mass and redshift (solid lines).}
\label{fig:seds}
\end{figure*}

To bolster the cartoon view presented in Figure~\ref{fig:demo}, we 
additionally consider the composite SEDs of normal star-forming galaxies in bins of stellar mass.
Due to the increasing levels of dust attenuation, we see a clear evolution of the composite SEDs  
in Figure~\ref{fig:seds}.  On average, the most massive star-forming galaxies
have characteristically dusty spectral shapes, with a 2175\AA\ dust feature
evident, whereas lower stellar mass galaxies
have decreasing amounts of dust obscuration.  We note that we see similar trends at higher and lower redshifts,
but are unable to make robust statements due to incompleteness.

We speculate that the small fraction of dusty, blue galaxies are mainly starbursts, 
as very few appear to be associated with X-ray sources.
Remarkably, we see that the spectral shapes of these galaxies are all very similar, irrespective of
stellar mass (upper right panel in Figure~\ref{fig:seds}).  This suggests that the same physical process
is dominating the stellar light, possibly a merger driven starburst.

While 28\% of galaxies with log(M$_{\star})>10$ have already quenched their star formation at $1<z<1.5$,
we find that 11\% may be in the process of shutting down star formation.  These galaxies have red colors
and low $\mathrm{L_{IR}/L_{UV}}$ ratios, occupying the lower envelope of the star formation sequence.
We compare the composite SEDs of these red, low-dust star-forming galaxies 
to that of quiescent galaxies at the same stellar mass and redshift (solid lines in
bottom right panel of Figure~\ref{fig:seds}). These galaxies have similar rest-frame
$V$--$J$ colors to quiescent galaxies, but somewhat bluer rest-frame $U$--$V$ colors,
consistent with the idea that they are in the process of shutting down star
formation and may soon migrate to the red sequence.

By studying four distinct populations of galaxies selected 
from the NMBS, we have demonstrated that quantifying the observed properties of the star 
formation sequence and how the sequence evolves with time requires a
thorough understanding of the selection techniques and biases.  
A consequence of the strong dependence of dust attenuation on stellar mass is that measurements of the star 
formation sequence will depend critically on the sample selection.
The gradual evolution we measure of the slope of the star formation sequence toward
shallower values at high-$z$ is driven by the combination of the curvature of the ``normal'' star
formation sequence and the evolution of the mass-completeness limits with redshift.  
An in-depth analysis of the physical properties of these galaxies and comparisons between the 
observations and models will help constrain the physical mechanism driving this potential curvature and the 
outliers from the star formation sequence.

\begin{acknowledgements}
We thank the NMBS collaboration for their contribution to this work, and 
the COSMOS and AEGIS teams for the release of high quality
multi-wavelength data sets to the community.
Support from NSF grant AST-0807974 and NASA grant NNX11AB08G is gratefully acknowledged.
\end{acknowledgements}

\facility{\emph{facilities}: Mayall (NEWFIRM)}

\addcontentsline{toc}{chapter}{\numberline {}{\sc References}}


\begin{thebibliography}{50}
\expandafter\ifx\csname natexlab\endcsname\relax\def\natexlab#1{#1}\fi

\bibitem[{{Birnboim} \& {Dekel}(2003)}]{Birnboim03}
{Birnboim}, Y., \& {Dekel}, A. 2003, \mnras, 345, 349

\bibitem[{{Bouch{\'e}} {et~al.}(2010){Bouch{\'e}}, {Dekel}, {Genzel}, {Genel},
  {Cresci}, {F{\"o}rster Schreiber}, {Shapiro}, {Davies}, \&
  {Tacconi}}]{Bouche10}
{Bouch{\'e}}, N., {Dekel}, A., {Genzel}, R., {et~al.} 2010, \apj, 718, 1001

\bibitem[{{Brammer} {et~al.}(2011){Brammer}, {Whitaker}, {van Dokkum},
  {Marchesini}, {Franx}, {Kriek}, {Labb{\'e}}, {Lee}, {Muzzin}, {Quadri},
  {Rudnick}, \& {Williams}}]{Brammer11}
{Brammer}, G.~B., {Whitaker}, K.~E., {van Dokkum}, P.~G., {et~al.} 2011, \apj,
  739, 24

\bibitem[{{Brammer} {et~al.}(2009){Brammer}, {Whitaker}, {van Dokkum},
  {Marchesini}, {Labb{\'e}}, {Franx}, {Kriek}, {Quadri}, {Illingworth}, {Lee},
  {Muzzin}, \& {Rudnick}}]{Brammer09}
---. 2009, \apjl, 706, L173

\bibitem[{{Brinchmann} {et~al.}(2004){Brinchmann}, {Charlot}, {White},
  {Tremonti}, {Kauffmann}, {Heckman}, \& {Brinkmann}}]{Brinchmann04}
{Brinchmann}, J., {Charlot}, S., {White}, S.~D.~M., {et~al.} 2004, \mnras, 351,
  1151

\bibitem[{{Bruzual} \& {Charlot}(2003)}]{BC03}
{Bruzual}, G., \& {Charlot}, S. 2003, \mnras, 344, 1000

\bibitem[{{Calzetti} {et~al.}(2000){Calzetti}, {Armus}, {Bohlin}, {Kinney},
  {Koornneef}, \& {Storchi-Bergmann}}]{Calzetti00}
{Calzetti}, D., {Armus}, L., {Bohlin}, R.~C., {et~al.} 2000, \apj, 533, 682

\bibitem[{{Chabrier}(2003)}]{Chabrier}
{Chabrier}, G. 2003, \pasp, 115, 763

\bibitem[{{Daddi} {et~al.}(2010){Daddi}, {Bournaud}, {Walter}, {Dannerbauer},
  {Carilli}, {Dickinson}, {Elbaz}, {Morrison}, {Riechers}, {Onodera}, {Salmi},
  {Krips}, \& {Stern}}]{Daddi10}
{Daddi}, E., {Bournaud}, F., {Walter}, F., {et~al.} 2010, \apj, 713, 686

\bibitem[{{Daddi} {et~al.}(2008){Daddi}, {Dannerbauer}, {Elbaz}, {Dickinson},
  {Morrison}, {Stern}, \& {Ravindranath}}]{Daddi08}
{Daddi}, E., {Dannerbauer}, H., {Elbaz}, D., {et~al.} 2008, \apjl, 673, L21

\bibitem[{{Daddi} {et~al.}(2007){Daddi}, {Dickinson}, {Morrison}, {Chary},
  {Cimatti}, {Elbaz}, {Frayer}, {Renzini}, {Pope}, {Alexander}, {Bauer},
  {Giavalisco}, {Huynh}, {Kurk}, \& {Mignoli}}]{Daddi07}
{Daddi}, E., {Dickinson}, M., {Morrison}, G., {et~al.} 2007, \apj, 670, 156

\bibitem[{{Dale} \& {Helou}(2002)}]{DH02}
{Dale}, D.~A., \& {Helou}, G. 2002, \apj, 576, 159

\bibitem[{{Damen} {et~al.}(2009){Damen}, {Labb{\'e}}, {Franx}, {van Dokkum},
  {Taylor}, \& {Gawiser}}]{Damen09}
{Damen}, M., {Labb{\'e}}, I., {Franx}, M., {et~al.} 2009, \apj, 690, 937

\bibitem[{{Dekel} {et~al.}(2009){Dekel}, {Birnboim}, {Engel}, {Freundlich},
  {Goerdt}, {Mumcuoglu}, {Neistein}, {Pichon}, {Teyssier}, \&
  {Zinger}}]{Dekel09}
{Dekel}, A., {Birnboim}, Y., {Engel}, G., {et~al.} 2009, \nat, 457, 451

\bibitem[{{Dutton} {et~al.}(2010){Dutton}, {van den Bosch}, \&
  {Dekel}}]{Dutton10}
{Dutton}, A.~A., {van den Bosch}, F.~C., \& {Dekel}, A. 2010, \mnras, 405, 1690

\bibitem[{{Elbaz} {et~al.}(2007){Elbaz}, {Daddi}, {Le Borgne}, {Dickinson},
  {Alexander}, {Chary}, {Starck}, {Brandt}, {Kitzbichler}, {MacDonald},
  {Nonino}, {Popesso}, {Stern}, \& {Vanzella}}]{Elbaz07}
{Elbaz}, D., {Daddi}, E., {Le Borgne}, D., {et~al.} 2007, \aap, 468, 33

\bibitem[{{Elvis} {et~al.}(2009){Elvis}, {Civano}, {Vignali}, {Puccetti},
  {Fiore}, {Cappelluti}, {Aldcroft}, {Fruscione}, {Zamorani}, {Comastri},
  {Brusa}, {Gilli}, {Miyaji}, {Damiani}, {Koekemoer}, {Finoguenov}, {Brunner},
  {Urry}, {Silverman}, {Mainieri}, {Hasinger}, {Griffiths}, {Carollo}, {Hao},
  {Guzzo}, {Blain}, {Calzetti}, {Carilli}, {Capak}, {Ettori}, {Fabbiano},
  {Impey}, {Lilly}, {Mobasher}, {Rich}, {Salvato}, {Sanders}, {Schinnerer},
  {Scoville}, {Shopbell}, {Taylor}, {Taniguchi}, \& {Volonteri}}]{Elvis09}
{Elvis}, M., {Civano}, F., {Vignali}, C., {et~al.} 2009, \apjs, 184, 158

\bibitem[{{Franx} {et~al.}(2008){Franx}, {van Dokkum}, {Schreiber}, {Wuyts},
  {Labb{\'e}}, \& {Toft}}]{Franx08}
{Franx}, M., {van Dokkum}, P.~G., {Schreiber}, N.~M.~F., {et~al.} 2008, \apj,
  688, 770

\bibitem[{{Gonz{\'a}lez} {et~al.}(2010){Gonz{\'a}lez}, {Labb{\'e}}, {Bouwens},
  {Illingworth}, {Franx}, {Kriek}, \& {Brammer}}]{Gonzalez10}
{Gonz{\'a}lez}, V., {Labb{\'e}}, I., {Bouwens}, R.~J., {et~al.} 2010, \apj,
  713, 115

\bibitem[{{Hopkins} \& {Beacom}(2006)}]{Hopkins06}
{Hopkins}, A.~M., \& {Beacom}, J.~F. 2006, \apj, 651, 142

\bibitem[{{Hwang} {et~al.}(2010){Hwang}, {Elbaz}, {Magdis}, {Daddi},
  {Symeonidis}, {Altieri}, {Amblard}, {Andreani}, {Arumugam}, {Auld}, {Aussel},
  {Babbedge}, {Berta}, {Blain}, {Bock}, {Bongiovanni}, {Boselli}, {Buat},
  {Burgarella}, {Castro-Rodr{\'{\i}}guez}, {Cava}, {Cepa}, {Chanial}, {Chapin},
  {Chary}, {Cimatti}, {Clements}, {Conley}, {Conversi}, {Cooray},
  {Dannerbauer}, {Dickinson}, {Dominguez}, {Dowell}, {Dunlop}, {Dwek}, {Eales},
  {Farrah}, {Schreiber}, {Fox}, {Franceschini}, {Gear}, {Genzel}, {Glenn},
  {Griffin}, {Gruppioni}, {Halpern}, {Hatziminaoglou}, {Ibar}, {Isaak},
  {Ivison}, {Jeong}, {Lagache}, {Le Borgne}, {Le Floc'h}, {Lee}, {Lee}, {Lee},
  {Levenson}, {Lu}, {Lutz}, {Madden}, {Maffei}, {Magnelli}, {Mainetti},
  {Maiolino}, {Marchetti}, {Mortier}, {Nguyen}, {Nordon}, {O'Halloran},
  {Okumura}, {Oliver}, {Omont}, {Page}, {Panuzzo}, {Papageorgiou}, {Pearson},
  {P{\'e}rez-Fournon}, {Garc{\'{\i}}a}, {Poglitsch}, {Pohlen}, {Popesso},
  {Pozzi}, {Rawlings}, {Rigopoulou}, {Riguccini}, {Rizzo}, {Rodighiero},
  {Roseboom}, {Rowan-Robinson}, {Saintonge}, {Portal}, {Santini}, {Sauvage},
  {Schulz}, {Scott}, {Seymour}, {Shao}, {Shupe}, {Smith}, {Stevens}, {Sturm},
  {Tacconi}, {Trichas}, {Tugwell}, {Vaccari}, {Valtchanov}, {Vieira},
  {Vigroux}, {Wang}, {Ward}, {Wright}, {Xu}, \& {Zemcov}}]{Hwang10}
{Hwang}, H.~S., {Elbaz}, D., {Magdis}, G., {et~al.} 2010, \mnras, 409, 75

\bibitem[{{Ilbert} {et~al.}(2009){Ilbert}, {Capak}, {Salvato}, {Aussel},
  {McCracken}, {Sanders}, {Scoville}, {Kartaltepe}, {Arnouts}, {Le Floc'h},
  {Mobasher}, {Taniguchi}, {Lamareille}, {Leauthaud}, {Sasaki}, {Thompson},
  {Zamojski}, {Zamorani}, {Bardelli}, {Bolzonella}, {Bongiorno}, {Brusa},
  {Caputi}, {Carollo}, {Contini}, {Cook}, {Coppa}, {Cucciati}, {de la Torre},
  {de Ravel}, {Franzetti}, {Garilli}, {Hasinger}, {Iovino}, {Kampczyk},
  {Kneib}, {Knobel}, {Kovac}, {Le Borgne}, {Le Brun}, {F{\`e}vre}, {Lilly},
  {Looper}, {Maier}, {Mainieri}, {Mellier}, {Mignoli}, {Murayama}, {Pell{\`o}},
  {Peng}, {P{\'e}rez-Montero}, {Renzini}, {Ricciardelli}, {Schiminovich},
  {Scodeggio}, {Shioya}, {Silverman}, {Surace}, {Tanaka}, {Tasca}, {Tresse},
  {Vergani}, \& {Zucca}}]{Ilbert09}
{Ilbert}, O., {Capak}, P., {Salvato}, M., {et~al.} 2009, \apj, 690, 1236

\bibitem[{{Karim} {et~al.}(2011){Karim}, {Schinnerer},
  {Mart{\'{\i}}nez-Sansigre}, {Sargent}, {van der Wel}, {Rix}, {Ilbert},
  {Smol{\v c}i{\'c}}, {Carilli}, {Pannella}, {Koekemoer}, {Bell}, \&
  {Salvato}}]{Karim11}
{Karim}, A., {Schinnerer}, E., {Mart{\'{\i}}nez-Sansigre}, A., {et~al.} 2011,
  \apj, 730, 61

\bibitem[{{Kennicutt}(1998)}]{Kennicutt98}
{Kennicutt}, Jr., R.~C. 1998, \araa, 36, 189

\bibitem[{{Kriek} {et~al.}(2009){Kriek}, {van Dokkum}, {Labb{\'e}}, {Franx},
  {Illingworth}, {Marchesini}, \& {Quadri}}]{Kriek09a}
{Kriek}, M., {van Dokkum}, P.~G., {Labb{\'e}}, I., {et~al.} 2009, \apj, 700,
  221

\bibitem[{{Labb{\'e}} {et~al.}(2005){Labb{\'e}}, {Huang}, {Franx}, {Rudnick},
  {Barmby}, {Daddi}, {van Dokkum}, {Fazio}, {Schreiber}, {Moorwood}, {Rix},
  {R{\"o}ttgering}, {Trujillo}, \& {van der Werf}}]{Labbe05}
{Labb{\'e}}, I., {Huang}, J., {Franx}, M., {et~al.} 2005, \apjl, 624, L81

\bibitem[{{Laird} {et~al.}(2009){Laird}, {Nandra}, {Georgakakis}, {Aird},
  {Barmby}, {Conselice}, {Coil}, {Davis}, {Faber}, {Fazio}, {Guhathakurta},
  {Koo}, {Sarajedini}, \& {Willmer}}]{Laird09}
{Laird}, E.~S., {Nandra}, K., {Georgakakis}, A., {et~al.} 2009, \apjs, 180, 102

\bibitem[{{Madau} {et~al.}(1996){Madau}, {Ferguson}, {Dickinson}, {Giavalisco},
  {Steidel}, \& {Fruchter}}]{Madau96}
{Madau}, P., {Ferguson}, H.~C., {Dickinson}, M.~E., {et~al.} 1996, \mnras, 283,
  1388

\bibitem[{{Magdis} {et~al.}(2010){Magdis}, {Rigopoulou}, {Huang}, \&
  {Fazio}}]{Magdis10}
{Magdis}, G.~E., {Rigopoulou}, D., {Huang}, J.-S., {et~al.} 2010, \mnras, 401,
  1521

\bibitem[{{Marcillac} {et~al.}(2006){Marcillac}, {Elbaz}, {Chary}, {Dickinson},
  {Galliano}, \& {Morrison}}]{Marcillac06}
{Marcillac}, D., {Elbaz}, D., {Chary}, R.~R., {et~al.} 2006, \aap, 451, 57

\bibitem[{{Muzzin} {et~al.}(2010){Muzzin}, {van Dokkum}, {Kriek}, {Labb{\'e}},
  {Cury}, {Marchesini}, \& {Franx}}]{Muzzin10}
{Muzzin}, A., {van Dokkum}, P., {Kriek}, M., {et~al.} 2010, \apj, 725, 742

\bibitem[{{Narayanan} \& {Dav{\'e}}(2012)}]{Narayanan12}
{Narayanan}, D., \& {Dav{\'e}}, R. 2012, ArXiv e-prints

\bibitem[{{Noeske} {et~al.}(2007){Noeske}, {Weiner}, {Faber}, {Papovich},
  {Koo}, {Somerville}, {Bundy}, {Conselice}, {Newman}, {Schiminovich}, {Le
  Floc'h}, {Coil}, {Rieke}, {Lotz}, {Primack}, {Barmby}, {Cooper}, {Davis},
  {Ellis}, {Fazio}, {Guhathakurta}, {Huang}, {Kassin}, {Martin}, {Phillips},
  {Rich}, {Small}, {Willmer}, \& {Wilson}}]{Noeske07a}
{Noeske}, K.~G., {Weiner}, B.~J., {Faber}, S.~M., {et~al.} 2007, \apjl, 660,
  L43

\bibitem[{{Pannella} {et~al.}(2009){Pannella}, {Carilli}, {Daddi}, {McCracken},
  {Owen}, {Renzini}, {Strazzullo}, {Civano}, {Koekemoer}, {Schinnerer},
  {Scoville}, {Smol{\v c}i{\'c}}, {Taniguchi}, {Aussel}, {Kneib}, {Ilbert},
  {Mellier}, {Salvato}, {Thompson}, \& {Willott}}]{Pannella09}
{Pannella}, M., {Carilli}, C.~L., {Daddi}, E., {et~al.} 2009, \apjl, 698, L116

\bibitem[{{Peng} {et~al.}(2010){Peng}, {Lilly}, {Kova{\v c}}, {Bolzonella},
  {Pozzetti}, {Renzini}, {Zamorani}, {Ilbert}, {Knobel}, {Iovino}, {Maier},
  {Cucciati}, {Tasca}, {Carollo}, {Silverman}, {Kampczyk}, {de Ravel},
  {Sanders}, {Scoville}, {Contini}, {Mainieri}, {Scodeggio}, {Kneib}, {Le
  F{\`e}vre}, {Bardelli}, {Bongiorno}, {Caputi}, {Coppa}, {de la Torre},
  {Franzetti}, {Garilli}, {Lamareille}, {Le Borgne}, {Le Brun}, {Mignoli},
  {Perez Montero}, {Pello}, {Ricciardelli}, {Tanaka}, {Tresse}, {Vergani},
  {Welikala}, {Zucca}, {Oesch}, {Abbas}, {Barnes}, {Bordoloi}, {Bottini},
  {Cappi}, {Cassata}, {Cimatti}, {Fumana}, {Hasinger}, {Koekemoer},
  {Leauthaud}, {Maccagni}, {Marinoni}, {McCracken}, {Memeo}, {Meneux}, {Nair},
  {Porciani}, {Presotto}, \& {Scaramella}}]{Peng10}
{Peng}, Y.-j., {Lilly}, S.~J., {Kova{\v c}}, K., {et~al.} 2010, \apj, 721, 193

\bibitem[{{Reddy} {et~al.}(2012){Reddy}, {Dickinson}, {Elbaz}, {Morrison},
  {Giavalisco}, {Ivison}, {Papovich}, {Scott}, {Buat}, {Burgarella},
  {Charmandaris}, {Daddi}, {Magdis}, {Murphy}, {Altieri}, {Aussel},
  {Dannerbauer}, {Dasyra}, {Hwang}, {Kartaltepe}, {Leiton}, {Magnelli}, \&
  {Popesso}}]{Reddy12}
{Reddy}, N., {Dickinson}, M., {Elbaz}, D., {et~al.} 2012, \apj, 744, 154

\bibitem[{{Reddy} {et~al.}(2010){Reddy}, {Erb}, {Pettini}, {Steidel}, \&
  {Shapley}}]{Reddy10}
{Reddy}, N.~A., {Erb}, D.~K., {Pettini}, M., {et~al.} 2010, \apj, 712, 1070

\bibitem[{{Reddy} {et~al.}(2006){Reddy}, {Steidel}, {Fadda}, {Yan}, {Pettini},
  {Shapley}, {Erb}, \& {Adelberger}}]{Reddy06}
{Reddy}, N.~A., {Steidel}, C.~C., {Fadda}, D., {et~al.} 2006, \apj, 644, 792

\bibitem[{{Sanders} {et~al.}(2007){Sanders}, {Salvato}, {Aussel}, {Ilbert},
  {Scoville}, {Surace}, {Frayer}, {Sheth}, {Helou}, {Brooke}, {Bhattacharya},
  {Yan}, {Kartaltepe}, {Barnes}, {Blain}, {Calzetti}, {Capak}, {Carilli},
  {Carollo}, {Comastri}, {Daddi}, {Ellis}, {Elvis}, {Fall}, {Franceschini},
  {Giavalisco}, {Hasinger}, {Impey}, {Koekemoer}, {Le F{\`e}vre}, {Lilly},
  {Liu}, {McCracken}, {Mobasher}, {Renzini}, {Rich}, {Schinnerer}, {Shopbell},
  {Taniguchi}, {Thompson}, {Urry}, \& {Williams}}]{Sanders07}
{Sanders}, D.~B., {Salvato}, M., {Aussel}, H., {et~al.} 2007, \apjs, 172, 86

\bibitem[{{Santini} {et~al.}(2012){Santini}, {Rosario}, {Shao}, {Lutz},
  {Maiolino}, {Alexander}, {Altieri}, {Andreani}, {Aussel}, {Bauer}, {Berta},
  {Bongiovanni}, {Brandt}, {Brusa}, {Cepa}, {Cimatti}, {Daddi}, {Elbaz},
  {Fontana}, {F{\"o}rster Schreiber}, {Genzel}, {Grazian}, {Le Floc'h},
  {Magnelli}, {Mainieri}, {Nordon}, {P{\'e}rez Garcia}, {Poglitsch}, {Popesso},
  {Pozzi}, {Riguccini}, {Rodighiero}, {Salvato}, {Sanchez-Portal}, {Sturm},
  {Tacconi}, {Valtchanov}, \& {Wuyts}}]{Santini12}
{Santini}, P., {Rosario}, D.~J., {Shao}, L., {et~al.} 2012, \aap, 540, A109

\bibitem[{{Tacconi} {et~al.}(2010){Tacconi}, {Genzel}, {Neri}, {Cox}, {Cooper},
  {Shapiro}, {Bolatto}, {Bouch{\'e}}, {Bournaud}, {Burkert}, {Combes},
  {Comerford}, {Davis}, {Schreiber}, {Garcia-Burillo}, {Gracia-Carpio}, {Lutz},
  {Naab}, {Omont}, {Shapley}, {Sternberg}, \& {Weiner}}]{Tacconi10}
{Tacconi}, L.~J., {Genzel}, R., {Neri}, R., {et~al.} 2010, \nat, 463, 781

\bibitem[{{Whitaker} {et~al.}(2011){Whitaker}, {Labb{\'e}}, {van Dokkum},
  {Brammer}, {Kriek}, {Marchesini}, {Quadri}, {Franx}, {Muzzin}, {Williams},
  {Bezanson}, {Illingworth}, {Lee}, {Lundgren}, {Nelson}, {Rudnick}, {Tal}, \&
  {Wake}}]{Whitaker11}
{Whitaker}, K.~E., {Labb{\'e}}, I., {van Dokkum}, P.~G., {et~al.} 2011, \apj,
  735, 86

\bibitem[{{Whitaker} {et~al.}(2010){Whitaker}, {van Dokkum}, {Brammer},
  {Kriek}, {Franx}, {Labb{\'e}}, {Marchesini}, {Quadri}, {Bezanson},
  {Illingworth}, {Lee}, {Muzzin}, {Rudnick}, \& {Wake}}]{Whitaker10}
{Whitaker}, K.~E., {van Dokkum}, P.~G., {Brammer}, G., {et~al.} 2010, \apj,
  719, 1715

\bibitem[{{Williams} {et~al.}(2009){Williams}, {Quadri}, {Franx}, {van Dokkum},
  \& {Labb{\'e}}}]{Williams09}
{Williams}, R.~J., {Quadri}, R.~F., {Franx}, M., {et~al.} 2009, \apj, 691, 1879

\bibitem[{{Wuyts} {et~al.}(2012){Wuyts}, {Forster Schreiber}, {Genzel}, {Guo},
  {Barro}, {Bell}, {Dekel}, {Faber}, {Ferguson}, {Giavalisco}, {Grogin},
  {Hathi}, {Huang}, {Kocevski}, {Koekemoer}, {Koo}, {Lotz}, {Lutz}, {McGrath},
  {Newman}, {Rosario}, {Saintonge}, {Tacconi}, {Weiner}, \& {van der
  Wel}}]{Wuyts12}
{Wuyts}, S., {Forster Schreiber}, N.~M., {Genzel}, R., {et~al.} 2012, ArXiv
  e-prints

\bibitem[{{Wuyts} {et~al.}(2011{\natexlab{a}}){Wuyts}, {F{\"o}rster Schreiber},
  {Lutz}, {Nordon}, {Berta}, {Altieri}, {Andreani}, {Aussel}, {Bongiovanni},
  {Cepa}, {Cimatti}, {Daddi}, {Elbaz}, {Genzel}, {Koekemoer}, {Magnelli},
  {Maiolino}, {McGrath}, {P{\'e}rez Garc{\'{\i}}a}, {Poglitsch}, {Popesso},
  {Pozzi}, {Sanchez-Portal}, {Sturm}, {Tacconi}, \& {Valtchanov}}]{Wuyts11a}
{Wuyts}, S., {F{\"o}rster Schreiber}, N.~M., {Lutz}, D., {et~al.}
  2011{\natexlab{a}}, \apj, 738, 106

\bibitem[{{Wuyts} {et~al.}(2011{\natexlab{b}}){Wuyts}, {F{\"o}rster Schreiber},
  {van der Wel}, {Magnelli}, {Guo}, {Genzel}, {Lutz}, {Aussel}, {Barro},
  {Berta}, {Cava}, {Graci{\'a}-Carpio}, {Hathi}, {Huang}, {Kocevski},
  {Koekemoer}, {Lee}, {Le Floc'h}, {McGrath}, {Nordon}, {Popesso}, {Pozzi},
  {Riguccini}, {Rodighiero}, {Saintonge}, \& {Tacconi}}]{Wuyts11b}
{Wuyts}, S., {F{\"o}rster Schreiber}, N.~M., {van der Wel}, A., {et~al.}
  2011{\natexlab{b}}, \apj, 742, 96

\bibitem[{{Wuyts} {et~al.}(2007){Wuyts}, {Labb{\'e}}, {Franx}, {Rudnick}, {van
  Dokkum}, {Fazio}, {F{\"o}rster Schreiber}, {Huang}, {Moorwood}, {Rix},
  {R{\"o}ttgering}, \& {van der Werf}}]{Wuyts07}
{Wuyts}, S., {Labb{\'e}}, I., {Franx}, M., {et~al.} 2007, \apj, 655, 51

\bibitem[{{Wuyts} {et~al.}(2008){Wuyts}, {Labb{\'e}}, {Schreiber}, {Franx},
  {Rudnick}, {Brammer}, \& {van Dokkum}}]{Wuyts08}
{Wuyts}, S., {Labb{\'e}}, I., {Schreiber}, N.~M.~F., {et~al.} 2008, \apj, 682,
  985

\bibitem[{{Zheng} {et~al.}(2007){Zheng}, {Bell}, {Papovich}, {Wolf},
  {Meisenheimer}, {Rix}, {Rieke}, \& {Somerville}}]{Zheng07}
{Zheng}, X.~Z., {Bell}, E.~F., {Papovich}, C., {et~al.} 2007, \apjl, 661, L41

\end{thebibliography}
\end{document}